\documentclass[aps,prl,reprint,twocolumn,floatfix,superscriptaddress,longbibliography]{revtex4-2}

\usepackage{standalone}
\newcommand{\bea}{\begin{equation}\begin{aligned}}

\newcommand{\Kr}[2]{\delta_{#1}^{#2}}
\newcommand{\eea}{\end{aligned}\end{equation}}
\newcommand{\aaa}{\hat{a}}

\newcommand{\aaas}{\hat{a}^{2}}
\newcommand{\daaa}{\hat{a}^{\dagger}}
\newcommand{\daaas}{\hat{a}^{\dagger2}}
\usepackage{amsthm}
\usepackage{lipsum}
\usepackage{amsfonts, amsmath, amssymb}

\newcommand{\G}{G}

\newcommand{\mean}[1]{\left\langle #1 \right\rangle}

\newcommand{\calpha}{\alpha^\ast}
\newcommand{\calphas}{\alpha^{\ast2}}
\usepackage{graphicx}
\usepackage{dcolumn}
\usepackage{bm}
\usepackage{dsfont}
\usepackage[normalem]{ulem}
\usepackage{color}
\usepackage[utf8]{inputenc}
\usepackage[colorlinks,citecolor=blue,linkcolor=blue,urlcolor=blue]{hyperref}
\usepackage{tikz}
\usepackage{braket}
\usepackage{physics}
\usepackage{color}
\usepackage{soul}
\usepackage{mathtools}
\usepackage{float}
\usepackage{esvect}
\usepackage{subfigure}
\usepackage{stmaryrd} 
\usepackage{enumerate}
\usepackage{bbold}
\newcommand{\abss}[1]{|#1|^2}

\include{physics}

\makeatletter
\def\maketitle{
	\@author@finish
	\title@column\titleblock@produce
	\suppressfloats[t]}
\makeatother

\begin{document}



\title{Emergent Equilibrium in All-Optical Single Quantum-Trajectory Ising Machines}

\author{Jacopo Tosca}
\affiliation{Universit\'{e} Paris Cit\'e, CNRS, Mat\'{e}riaux et Ph\'{e}nom\`{e}nes Quantiques, 75013 Paris, France}
\author{Marcello Calvanese Strinati}
\affiliation{Centro Ricerche Enrico Fermi (CREF), Via Panisperna 89a, 00184 Rome, Italy}

\author{Claudio Conti}
\affiliation{Centro Ricerche Enrico Fermi (CREF), Via Panisperna 89a, 00184 Rome, Italy}
\affiliation{Physics Department, Sapienza University of Rome, 00185 Rome, Italy}
\author{ Cristiano Ciuti}
\affiliation{Universit\'{e} Paris Cit\'e, CNRS, Mat\'{e}riaux et Ph\'{e}nom\`{e}nes Quantiques, 75013 Paris, France}

\begin{abstract}
\noindent 
We investigate the dynamics of multi-mode optical systems driven by two-photon processes and subject to non-local losses, incorporating quantum noise at the Gaussian level. Our findings show that the statistics retrieved from a single Gaussian quantum trajectory exhibits emergent thermal equilibrium governed by an Ising Hamiltonian, encoded in the dissipative coupling between modes. The system's effective temperature is set by the driving strength relative to the oscillation threshold. Given the ultra-short time scales typical of all-optical devices, our study demonstrates that such multi-mode optical systems can operate as ultra-fast Boltzmann samplers, paving the way towards the realization of efficient hardware for combinatorial optimization, with promising applications in machine learning and beyond.
\end{abstract}
\date{\today}
\maketitle

{\it Introduction ---}
The Ising model describes a system of interacting binary spins with arbitrary spin-spin couplings and it is a fundamental cornerstone of statistical physics. The problem of minimizing the interaction energy notably belongs to the NP-hard complexity class using conventional computers~\cite{Barahona_1982}, as the time required to find its exact solution scales exponentially with the system size. Thanks to the possibility of efficiently casting the cost function of several hard optimization problems as specific Ising energies~\cite{10.3389/fphy.2014.00005}, the Ising model nowadays plays a pivotal role in diverse fields, such as artificial intelligence, machine learning, and bioinformatics (see~\cite{Mohseni2022} and references therein). Finding the optimal solution of the original optimization problem translates into minimizing the Ising energy. This circumstance has triggered a rapidly growing interest in analog physical hardware, known as Ising machines, designed to find low-energy solutions of large Ising models more efficiently than conventional computer architectures.

Ising machines have been proposed using several different physical systems~\cite{Byrnes_2011,Johnson2011,monroequantumsimulation2010,hgoto2019cim,bohm2019,gershenzonrapidlaser2019,PhysRevA.88.063853}. Among them, photonic systems hold great promise for surpassing integrated electronics in terms of data processing and sampling rates~\cite{Ambs2010}. The coherent Ising machine (CIM)~\cite{PhysRevA.88.063853,Yamamoto2020} addresses the Ising problem by encoding spins into the phases of optical pulses in a network of coupled optical parametric oscillators (OPOs). When driven above the oscillation threshold, the mode fields oscillate with a phase $0$ or $\pi$ with respect to the reference phase enforced by the pump~\cite{landauer1971}. These two states encode the ``up'' and ``down'' spin configurations of a classical Ising spin. Spin-spin couplings are realized through an electronic feedback loop, which calculates and applies interactions based on measured outputs. While the optical components operate at ultra-fast timescales, the overall speed of the CIM is constrained by the slower electronic feedback that governs the system’s evolution.

To overcome these issues, a multi-mode all-optical Ising machine, driven by parametric two-photon pumping and implementing dissipative mode coupling via spatial light modulators, was recently proposed in~\cite{Strinati2021} and studied at the mean-field level. In this context, the Ising spins are encoded into the phase of the parametrically generated field on different spatial points of the optical wavefront. Similarly to the CIM, the OPO amplitude dynamics behaves as a gradient descent in the energy landscape of coupled OPOs, tending to minimize a Lyapunov function that, in the regime of operation where all OPOs have equal amplitudes, encodes the Ising energy determined by the mode coupling~\cite{roychowdhury2022} specified by a symmetric coupling (graph) matrix $J_{ij}$.

In several practical situations, mean-field machines operate with unconstrained OPO amplitudes, and thus the Lyapunov function often significantly differs from the desired Ising energy~\cite{Yamamoto2020,roychowdhury2022}. This issue manifests itself in the presence of a plethora of stable solutions for the mean-field machine, which correspond to excited energy states of the simulated Ising model, effectively limiting the efficiency of the system as an Ising minimizer~\cite{PhysRevLett.126.143901,kalinin2020}. One possible way to circumvent this issue is by exploiting classical or quantum noise to let the system stochastically explore in time the spin configuration space~\cite{PhysRevResearch.4.013009}. Previous literature on electro-optical CIMs has shown that, when noise is sufficiently strong, the system samples low energy states of the Ising model from a thermal distribution, allowing the use of the CIM as a Boltzmann sampler~\cite{e18100365,bohmboltzmann2022}.

At a classical level, noise can be arbitrarily engineered to artificially induce thermal fluctuations. At quantum level instead, both the frequency and the amplitude of the noise are self-regulated by the quantum stochastic dynamics. To date, a systematic study on how quantum noise impacts the dynamics of the all-optical Ising machine is still missing, and consequently a method for quantifying the probability of success in the presence of noise has yet to be established.


In this letter, we theoretically investigate the quantum dynamics of multi-mode optical systems driven by two-photon processes and subject to non-local dissipation, analyzed at the Gaussian level by means of the quantum Gaussian trajectory approach~\cite{Wouter2019,Wouter2020}. By describing the quantum trajectories as the stochastic dynamics of the first and second order Gaussian moments, we demonstrate that one individual trajectory dynamically explores various spin configurations over time, statistically revealing an emergent thermal equilibrium governed by the target Ising Hamiltonian. This approach provides a pathway for determining the ground state of the Ising problem, effectively addressing the minimization challenge. 

\begin{figure}[t!]
\includegraphics[width = 1.0\hsize]{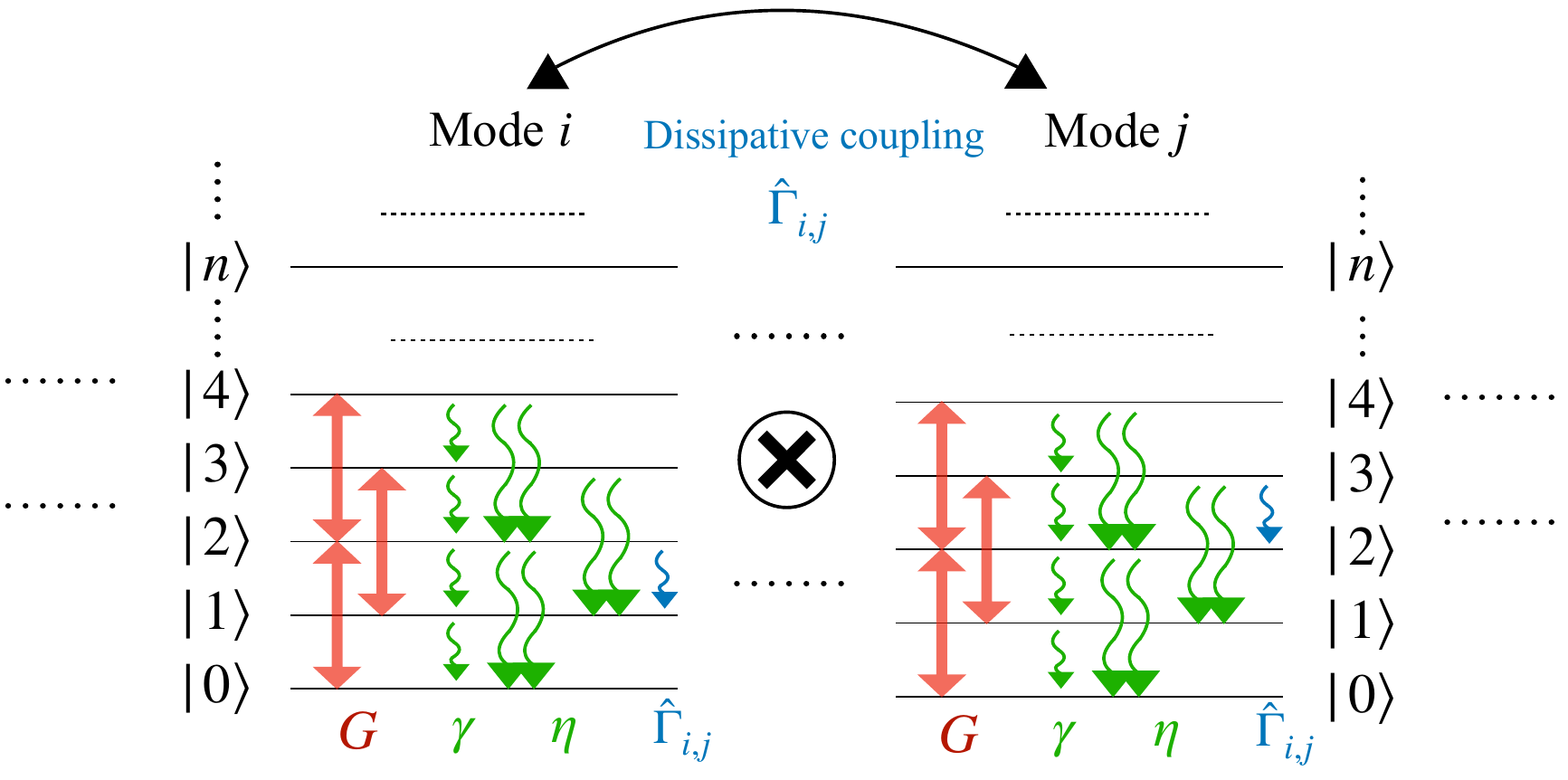}\\
\includegraphics[width = 1.0\hsize]{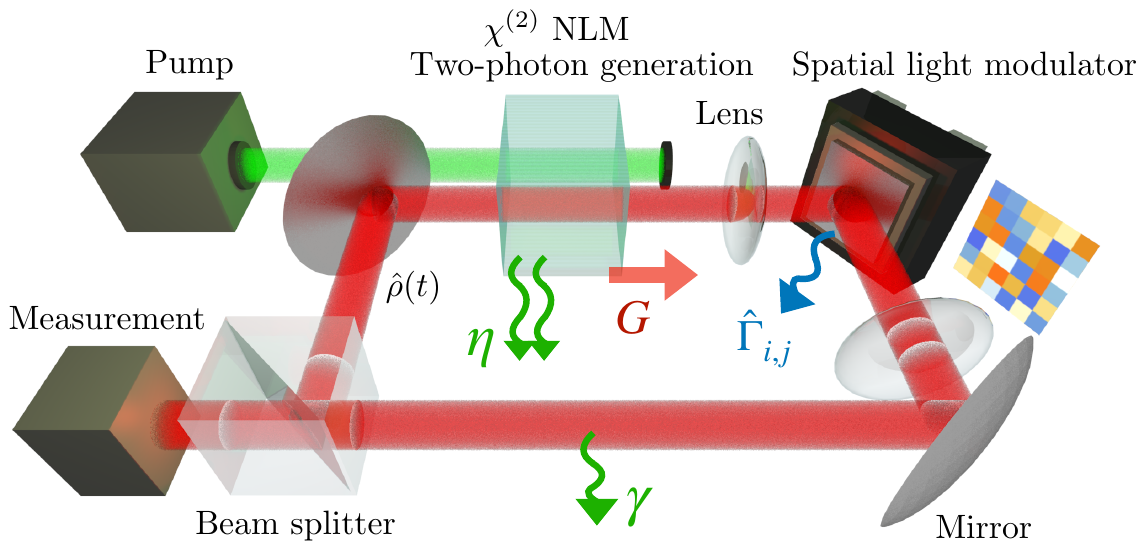}
  \caption{Top panel: Sketch of the considered system, describing $N$ optical modes driven by a two-photon pump with strength $G$. Each mode experiences single-photon losses at a rate $\gamma$ and two-photon losses at a rate $\eta$. Dissipative interactions between modes $i$ and $j$ are implemented via non-local jump operators $\hat \Gamma_{i,j}$, which is a function of the matrix $J_{ij}$ encoding the simulated Ising model (see text for the definitions).
Bottom panel: Example of a specific optical system that falls within the class described in the top panel.} 
\label{Fig:Schema_apparato}
\end{figure}
\begin{figure*}[t!]
\includegraphics[width = 0.325\hsize]{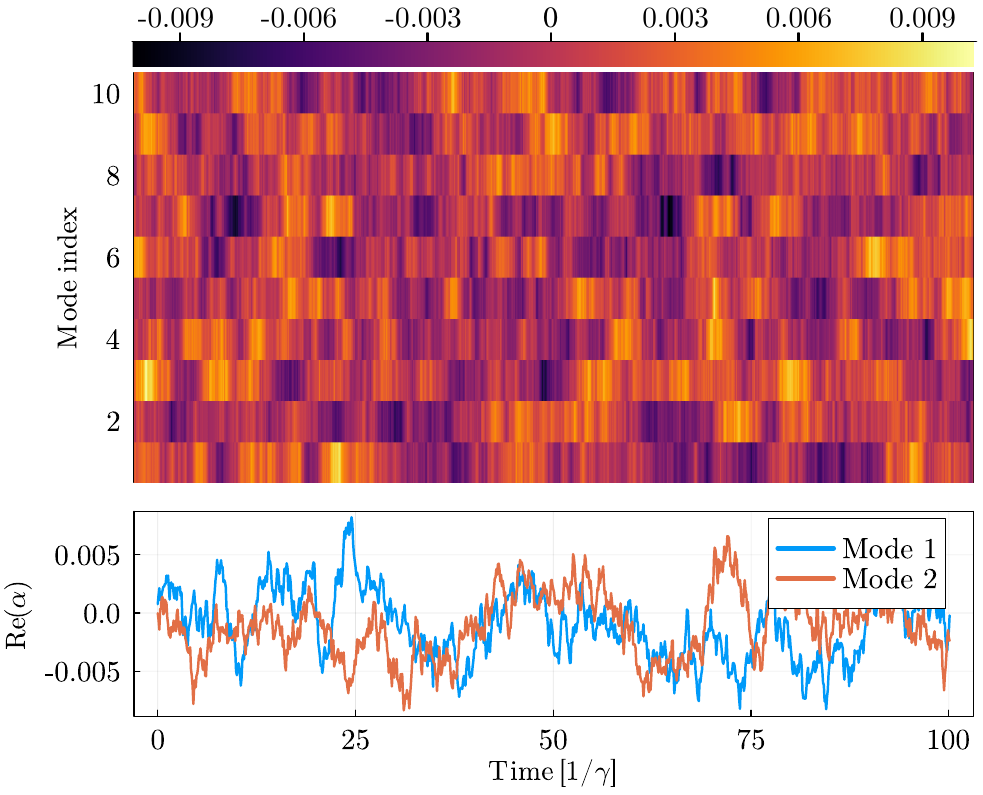}
\includegraphics[width = 0.325\hsize]{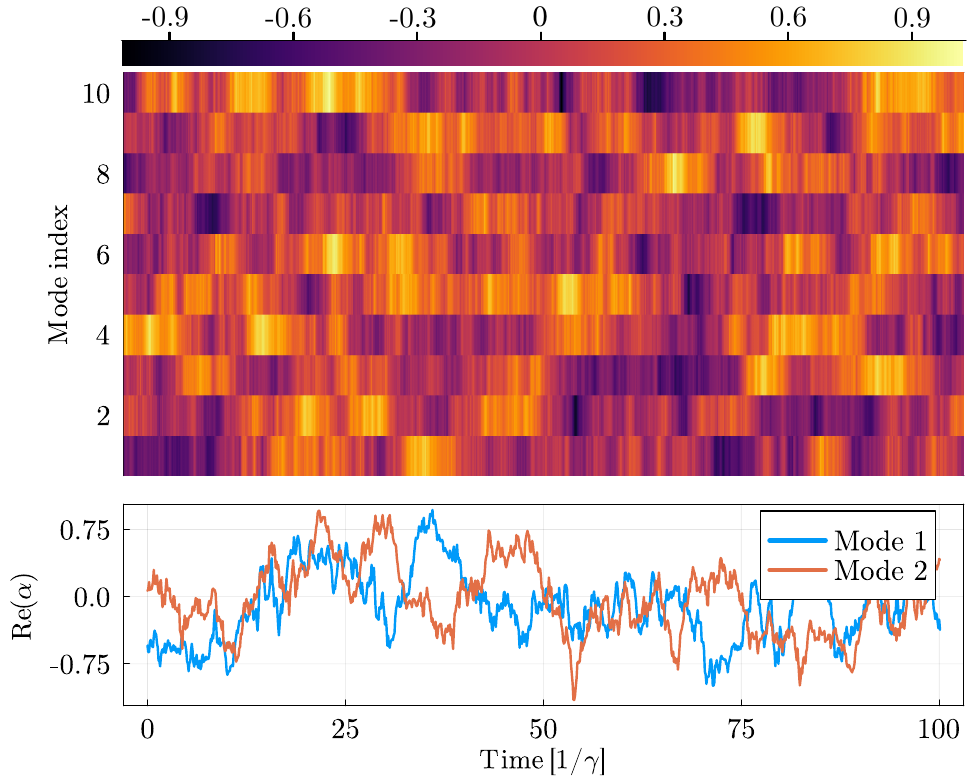}
\includegraphics[width = 0.325\hsize]{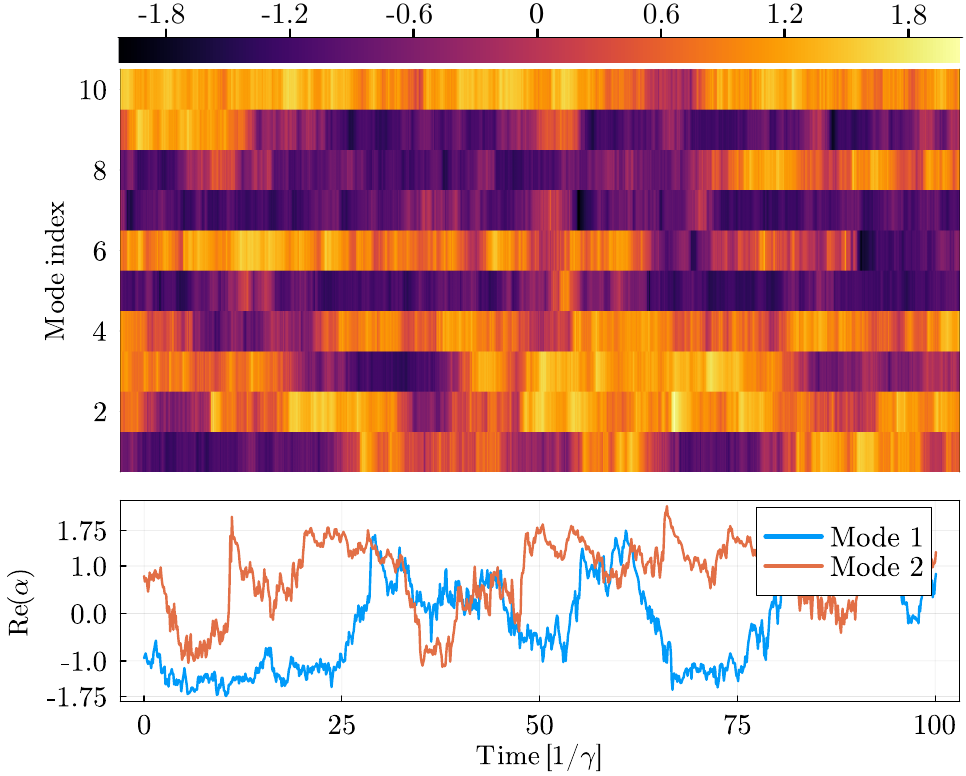}
\caption{Top panels: Heat map of the field quadratures $\Re(\alpha_i)$ for a SK graph with $N=10$ modes versus time (in units of $1/\gamma$) and the mode index $i$. Bottom panels: Illustrative curves of the temporal evolution for modes $1$ and $2$ to showcase the fluctuating pattern in different pump regimes $G/G_{\rm th}$ (other modes display the same qualitative behaviour). Left panels: $G/G_{\rm th} = 0.01$.  Central panels: $G/G_{\rm th} = 0.75$. Right panels: $G/G_{\rm th} = 1.5$. The two-photon   loss parameter is set to $\eta/\gamma = 0.1$.}
\label{mode_scan_0_75:Z}
\end{figure*}

{\it Theoretical framework ---}  
We consider a class of physical systems describing $N$ optical modes with the same frequency $\omega_0$, subject to coherent two-photon driving \cite{Rota2019}, but no Hamiltonian coupling between modes. The coupling is dissipative via non-local losses. An example of a concrete physical realization from~\cite{Strinati2021} is shown in the bottom panel of Fig.~\ref{Fig:Schema_apparato}.  In this case, two-photon driving is achieved by degenerate parametric down conversion, realized by pumping a second-order nonlinear medium ($\chi^{(2)}$ NLM) with a laser field at amplitude $G$ and frequency $2\omega_0$, where one pump photon is down-converted into two OPO photons. Non-local losses induce a dissipative coupling between the modes, which depends on $J_{ij}$ defining the simulated Ising model, as depicted in the top panel of Fig.~\ref{Fig:Schema_apparato}. Intrinsic cavity losses and nonlinear pump saturation give rise to additional one- and two-photon losses, at a rate $\gamma$ and $\eta$, respectively.

In the reference frame rotating at the mode frequency, for weak pump depletion and large pump mode intrinsic loss, two-photon driving is described by the following Hamiltonian~\cite{PhysRevA.43.6194}
\begin{equation}\label{Hamiltonian}
    \hat H = \frac{\mathrm{i}\hbar G}{2}\sum_{i=1}^{N}\left(\daaas_{i}-\aaas_{i}\right) \,\, ,
\end{equation}  
where $\hat a_i$ and $\hat a^\dag_i$ are the bosonic annihilation and creation operators of a OPO photon on mode $i$. Throughout this work, we consider real $G$, and describe the system dynamics by resorting to the formalism of Gaussian quantum trajectories~\cite{Petruccione,app8091427}. In this framework, the time evolution of the expectation value $\langle \hat{\mathrm{O}} \rangle$ of a generic observable on a single quantum trajectory, under heterodyne unraveling for all $n_{\rm dc}$ decay channels, is given by
\begin{eqnarray}
&&d\langle\hat{\mathrm{O}}\rangle= \frac{\mathrm{i}}{\hbar} \langle[\hat{H}, \hat{\mathrm{O}}\rangle] dt \nonumber \\
&&-\frac{1}{2} \sum_{s=1}^{n_{\rm dc}}\,\sum_{i=1}^N \sum_{j=1}^{i_s}\left(\left\langle\left\{\hat{\Gamma}^{(s)}_{i, j}{}^{\dagger} \hat{\Gamma}^{(s)}_{i, j} ,\hat{\mathrm{O}}\right\}\right\rangle-2\left\langle\hat{\Gamma}^{(s)}_{i, j}{}^{\dagger} \hat{\mathrm{O}} \hat{\Gamma}^{(s)}_{i, j}\right\rangle\right) d t \nonumber\\
&&+ \sum_{s=1}^{n_{\rm dc}}\,\sum_{i=1}^N \sum_{j=1}^{i_s}\left[\left\langle\hat{\Gamma}^{(s)}_{i, j}{}^{\dagger}(\hat{\mathrm{O}}-\langle\hat{\mathrm{O}}\rangle)\right\rangle d Z^{(s)}_{i, j}\right. \nonumber\\
&&\hspace{1.9cm}+\left.\left\langle(\hat{\mathrm{O}}-\langle\hat{\mathrm{O}}\rangle) \hat{\Gamma}^{(s)}_{i, j}\right\rangle d Z^{(s)}_{i, j}{}^*\right] \,\, ,
\label{eq:operatortimeevolution1}
\end{eqnarray}
where $i_1=i_2=N$ and $i_3=i-1$, and $dZ^{(s)}_{i,j} = (dW^{(s)}_{x,i,j} + \mathrm{i}\, dW^{(s)}_{p,i,j})/\sqrt{2}$ is a complex Wiener noise term satisfying $|d Z^{(s)}_{i,j}|^2 = dt$. 
In our system in Fig.~\ref{Fig:Schema_apparato}, the $n_{\rm dc}=3$ dissipation channels are described by the following jump operators: (i)~$\hat \Gamma^{(1)}_{i,j} = \sqrt{\gamma - \sum_{k=1}^{N} |J_{ik}|} \hat a_i \delta_{i,j}$ for one-photon loss, (ii)~$\hat \Gamma^{(2)}_{i,j} = \sqrt{\eta} \,{\hat a_i} ^2\,\delta_{i,j} $ for two-photon loss (nonlinear pump saturation), where $\delta_{i,j}$ is the Kr\"{o}necker delta, and (iii)~$\hat \Gamma^{(3)}_{i,j} = \sqrt{|J_{ij}|} (\hat a_{i} - J_{ij} / |J_{ij}| \ \hat a_{j})$ for the mode coupling, which comprises $N(N-1)/2$ loss terms. As detailed in the Supplementary Material (SM), the non-local loss terms  $\hat \Gamma^{(3)}_{i,j}$, combined with the one-photon loss terms $\hat \Gamma^{(1)}_{i,j}$, effectively encode a classical Ising Hamiltonian with coupling matrix ${J}_{ij}$~\cite{Goto2019}.

Hereafter, the heterodyne quantum trajectories in Eq.~\eqref{eq:operatortimeevolution1} are solved in the Gaussian approximation, which is expected to faithfully capture the relevant physics in the considered regime of loss parameters~\cite{Wouter2020}. To proceed with our numerical analysis, we consider two types of coupling matrices $J_{ij}$ defining fully-connected random graphs: (i) The spin-glass Sherrington-Kirkpatrick (SK) graph, where $J_{ij}$ are random numbers from a Gaussian distribution~\cite{PhysRevLett.35.1792}, and (ii) The complete random binary graph (K)~\cite{schneider1993graphs}, where $J_{ij}=\pm J$ with $J>0$ and sign randomly chosen with equal probability. The quantum state is fully determined by its first-order moments $\alpha_j=\mean{\aaa_j}$ and second-order moments $
    u_{ij}=\mean{\aaa_i\aaa_j}-\alpha_i\alpha_j$ and 
    $v_{ij}=\mean{\daaa_i\aaa_j}-\calpha_i\alpha_j$ (for more details, see Supplementary Material).
The time evolution of $\alpha_i=\langle\hat a_i\rangle$ from Eq.~\eqref{eq:operatortimeevolution1} for fixed initial conditions defines a single quantum trajectory for the OPO quadratures, from which the Ising spins state is retrieved.

Classically, at the mean-field level, when the pump $G$ is above the oscillation threshold value $G_{\rm th} = (\gamma - \lambda_{\text{max}})/2$ where $\lambda_{\text{max}}$ is the maximal eigenvalue of $J_{ij}$~\cite{PhysRevE.95.022118}, parametric amplification takes place. The dynamics amplifies the real part of the OPO amplitudes and suppresses their imaginary parts, making the OPOs eventually converge towards a real-amplitude steady state identified by the fixed points of the classical equations of motion describing the time evolution~\cite{PhysRevA.88.063853,PhysRevA.100.023835}. In contrast, in the presence of noise, a single quantum trajectory displays a strongly fluctuating pattern, as in bottom panels of Fig.~\ref{mode_scan_0_75:Z}. Due to parametric drive, the imaginary part of $\alpha_i(t)$ always fluctuates around zero. Instead, the real part of $\alpha_i(t)$ explores the quadrature space in a nontrivial, stochastic way that strongly depends on all system parameters.
At any time $t$, we define the Ising state ${\bm \sigma}(t)=(\sigma_1(t),\ldots,\sigma_N(t))$ from the mode real quadratures as $\sigma_i(t)={\rm sign}({\rm Re}[\alpha_i(t)])$. Accordingly, the classical Ising energy associated to ${\bm \sigma}(t)$ is computed as $E[{\bm \sigma}(t)] = -\frac{1}{2}\sum_{i,j=1}^{N} J_{ij} \sigma_i(t) \sigma_j(t)$.

Figure~\ref{mode_scan_0_75:Z} shows a prototypical individual quantum Gaussian trajectory for SK graph with $N=10$ modes. Top panels show heat maps of the dependence of $\Re(\alpha_i)$ for every mode index $i$ versus time, and bottom panels highlight the oscillating dynamics, for three different pump regimes: $G$ far below threshold (left), $25\%$ below threshold (center), $50\%$ above threshold (right). From each individual quantum trajectory, by monitoring how the spins $\sigma_i$ flip over time, we retrieve the time evolution of the Ising energy $E$. By simulating a single Gaussian quantum trajectory for a sufficiently long time $t_{\rm max}\gg\gamma^{-1}$ (we specifically use $t_{\rm max}=20,000\,\gamma^{-1}$), we can extract the probability distribution function, denoted by $P(E)$, defined as the number of times a given spin configuration at energy $E$ is found, divided by the total number of sampled times.

Figure~\ref{fig:2_graph_boltzmann} shows the recovered $P(E)$ in logarithmic scale, at a fixed pump amplitude, for two specific instances of SK (left panel) and K (right panel) graphs. The statistics for both graphs is well described by a Boltzmann distribution $P_B(E)=e^{-E/k_BT_{\rm eff}}/Z$ represented by the red solid line in the figure, where $Z$ is the partition function. For the SK graph, where $J_{ij}$ takes continuous values, all different spin configurations are associated to different energies, and thus the energy-multiplicity $n(E)=2$ of the spin configurations is uniform in $E$. Instead, for the K graph, the binary nature of $J_{ij}$ causes $n(E)$ to be non-uniform, strongly peaked around intermediate values of $E$ in the energy spectrum. This fact explains the qualitatively different distribution of the energy levels in the figure for the two graphs. The effective temperature $T_{\rm eff}$ is retrieved by fitting the numerical data of $P(E)$ with the Boltzmann distribution. Our numerical analysis demonstrates an emergent thermal equilibrium governed by the Ising energy, which favors lower-energy states over higher-energy ones in terms of statistical occurrences. The remarkable consequence is that the ground-state solution exponentially emerges as the most probable solution, provided that the physical evolution time is sufficiently long. 

Figure~\ref{fig:temperatura efficace} reports the dependence of the effective temperature $T_{\rm eff}$ on the driving strength $G$ for the two considered graphs. By simulating the single quantum trajectories up to $t_{\rm max}$ as in Fig.~\ref{fig:2_graph_boltzmann}, we observe that the effective temperature appears to be rather constant below threshold (the relevant scale is given by the loss rate $\gamma$), while it significantly decreases above threshold. The inset shows the fitted Boltzmann distributions in logarithmic scale for different values of $G/G_{\rm th}$. The data reveal that the fitted Boltzmann distributions for different values of $G/G_{\rm th}$ intersect within a narrow energy range. This fact notably implies that, by tuning the pump intensity, we can control the effective temperature $T_{{\rm eff}}$, further enhancing the probability of finding low-energy states to the left of the intersection region (in particular, the ground state) while reducing that of finding high-energy states, to the right of the intersection region. Note that emergent equilibrium has been evidenced in the many-body dynamics of bistable systems with optical Kerr nonlinearities~\cite{Foss-Feig2017}. Additionally, we recover the Boltzmann distribution {\it also} in the limit of vanishingly small pump amplitude $G\rightarrow0^+$, implying that hard driving is not required, as it has been recently evidenced for linear optical systems~\cite{Ramesh2024}. 

\begin{figure}[t!]
\includegraphics[width = 1.0\hsize]{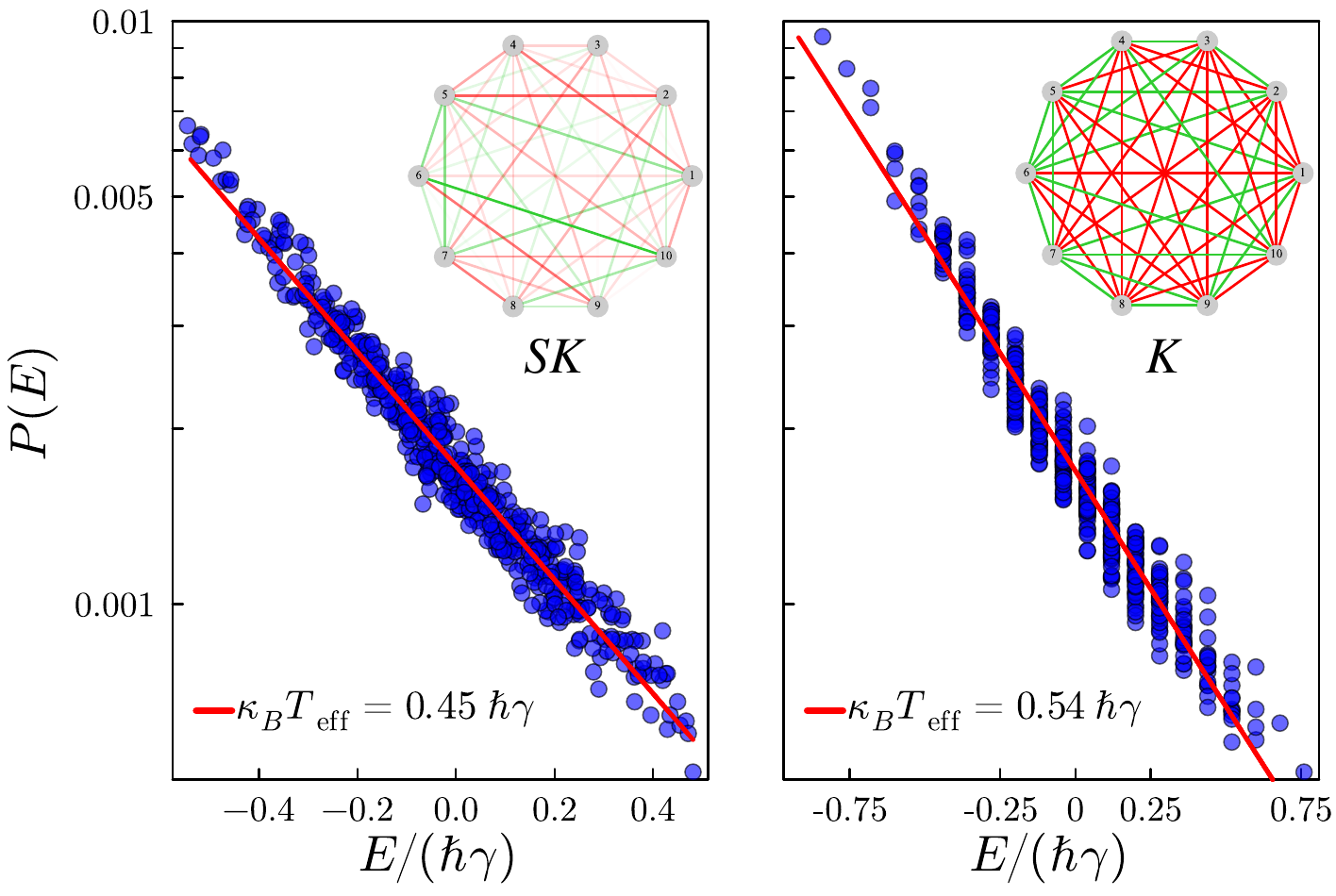}
\caption{Probability distribution function $P(E)$ in logarithmic scale as a function of the Ising energy $E$ for a system with $N = 10$ modes and $J_{ij}$ describing (Left) SK graph, and (Right) K graph. Graph connectivity is shown as green and red lines for positive and negative $J_{ij}$, respectively. The different intensity of the connectivity lines in left panel reflects the fact that $J_{ij}$ for the SK graph takes continuous values, while for the K graph $J_{ij}$ is binary. For both graphs, the scaling law $P(E) \propto e^{-E / (k_B T_{{\rm eff}})}$ is observed, revealing an emergent thermal equilibrium with an effective temperature $T_{\rm eff}$. Other parameters: $G/G_{\rm th} = 1.25$, $\eta/\gamma = 0.1$ and $200,000$ time samples taken from a single quantum trajectory evolved for a total time of $t_{\rm max}=20,000 \, \gamma^{-1}$.}
\label{fig:2_graph_boltzmann}
\end{figure}

\begin{figure}[t!]
\includegraphics[width = 1.0\hsize]{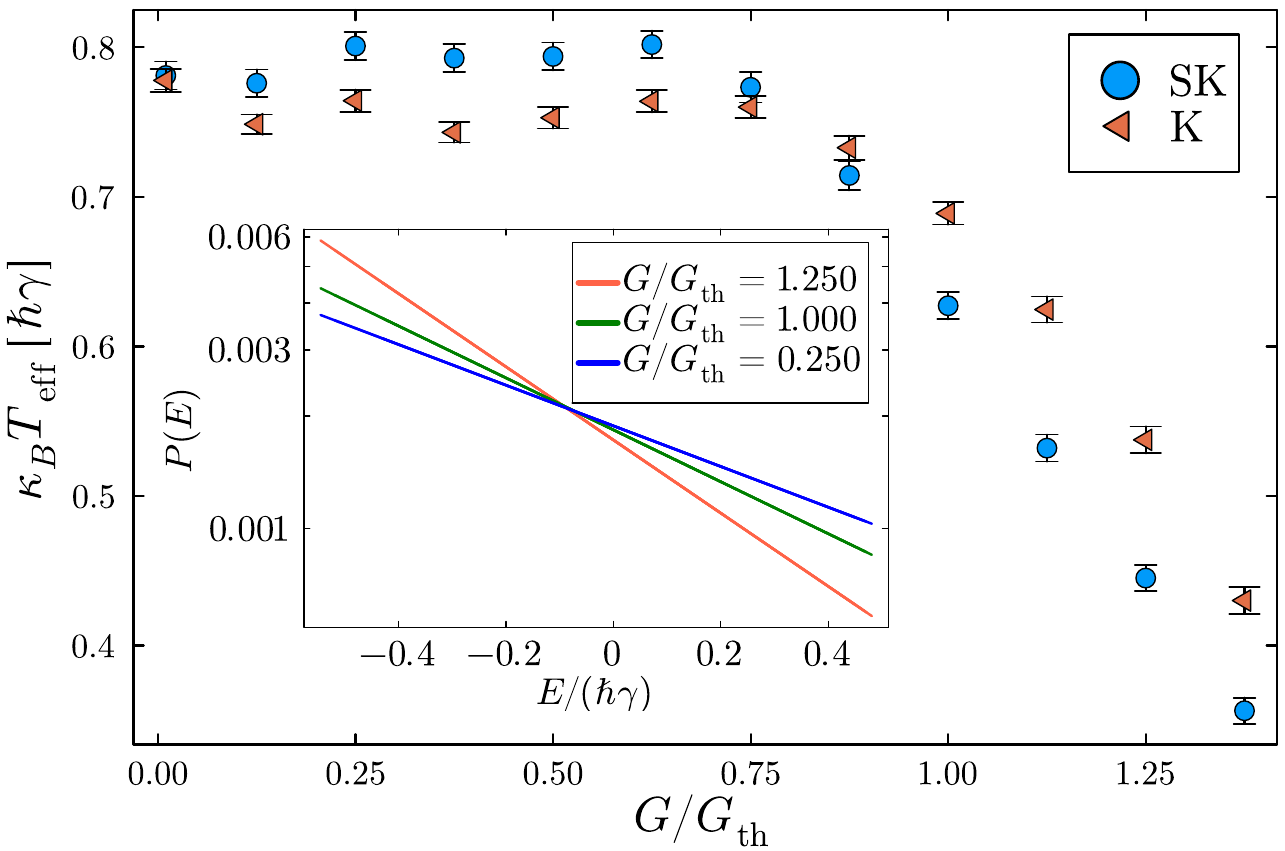}
  \caption{Effective temperature $T_{\rm eff}$ as a function of normalized pump amplitude $G/G_{\rm th}$ for the two different graphs with $N=~10$. Blue circles correspond to the SK graph, while the red triangles correspond to the K graph. The error bar represents one standard deviation, derived from fitting the Boltzmann distribution at various pump intensities $G/G_{\rm th}$. The inset shows the scaling law $P \propto e^{-E/k_BT_{\rm eff}}$ for different values of the pump intensity $G/G_{\rm th}$ for the SK graph. The fitted lines overlap within a narrow energy range. We use $\eta/\gamma = 0.1$, sampling $200,000$ times for each point ($t_{\rm max} = 20,000  \ \gamma^{-1}$).}
\label{fig:temperatura efficace}
\end{figure} 

Our findings demonstrate that the quantum noise acts as a resource in the all-optical Ising machine, enabling an exploration of encoded Ising spin configurations that gives rise to an emergent thermal equilibrium.  As the proposed implementation is all-optical, it allows for ultra-fast sampling of the spin states in an experimental setup.
This rapid sampling ensures that the ground-state probability exponentially surpasses that of other states. Additionally, by iterating the sampling procedure over time, confidence in the statistically predicted ground state can be continuously improved, in stark contrast with Ising machines operating in the mean-field limit. In this work we reported results with $N=10$ modes to clearly show the emergent thermal distribution over all Ising energy levels while ensuring reasonable computational times. Similar results have been however checked for larger values of $N$. As an all-optical system based on available technology can in principle encode from thousands to millions of spins~\cite{Strinati2021}, and autonomously (i.e., without feedback) perform the computation, our findings pave the way towards the realization of a Boltzmann sampler capable of delivering low-energy Ising solutions with high probability on a much shorter time scale compared to existing devices relying on electronics.

{\it Conclusions and outlooks ---}
We have demonstrated that multi-mode optical systems driven by two-photon processes and non-local losses reach thermal equilibrium governed by an Ising Hamiltonian. By analyzing quantum trajectories at the Gaussian level, we showed how quantum noise facilitates exploration of the Ising phase space and enables identification of the Ising ground state through emergent Boltzmann statistics, with the effective temperature controlled by the driving strength.

Our findings highlight the potential of all-optical Ising machines as ultra-fast Boltzmann samplers~\cite{gotoboltzmann2018}, promising advancements in optimization and machine learning~\cite{PhysRevX.8.021050,aaw1147}. These systems operate on ultrafast timescales, offering superior speed and scalability compared to conventional methods. Crucially, their autonomous design eliminates feedback loops, providing a practical route for implementing Ising solvers for complex coupling graphs.

Future work will extend these results to larger systems, investigate non-Gaussian noise, and validate the implementation experimentally, paving the way for leveraging quantum optical systems in solving optimization problems and studying spin systems.

We thank Zejian Li and V. Heyraud for help on the Gaussian trajectories.



%

\clearpage

\title{{\bf Supplementary Material for the article:}\\ ``Emergent Equilibrium in All-Optical Single Quantum-Trajectory Ising Machines"}
\setcounter{page}{1}
\setcounter{equation}{0}
\setcounter{figure}{0}
\renewcommand{\theequation}{S.\arabic{equation}}
\renewcommand{\thefigure}{S.\arabic{figure}}
\pagestyle{empty}
\date{\today}
\maketitle
\onecolumngrid

\section{Gaussian quantum trajectories}
Under the Gaussian approximation, the quantum trajectories are determined by the stochastic dynamics of the first and second order Gaussian moments \cite{Wouter2020}, namely: 
\begin{align}
    \alpha_j&=\mean{\aaa_j},\\
    u_{nm}&=\mean{\aaa_n\aaa_m}-\alpha_n\alpha_m,\\
    v_{nm}&=\mean{\daaa_n\aaa_m}-\calpha_n\alpha_m.
\end{align}
For our Hamiltonian and dissipators, described in the main text, such equations read:

\bea\label{Field_dynamics_equation}
   \frac{d\alpha_n}{dt} = &\big[ - \frac{\gamma}{2} \alpha_n + \G \calpha_n
   -  \eta\left(\abss{\alpha_n}\alpha_n + 2\alpha_{n}v_{nn}+\calpha_n u_{nn}\right) \big] + \frac{1}{2} \sum_{j} J_{jn} \alpha_j \\ &+\sum_{j}\sqrt{\gamma - \sum_i |J_{ij}|}\left( v_{jn}dZ_{j,j}^{(1)} + u_{jn}dZ_{j,j}^{(1)\ast}\right) + 2\sum_j \sqrt{\eta}\left( \calpha_j v_{jn} dZ_{j,j}^{(2)} + \alpha_j u_{jn} dZ_{j,j}^{(2)\ast} \right)  \\ 
   &+ \sum_{i}\sum_{j =1}^{i-1} \sqrt{|J_{ij}|} \Big( (v_{in} - \frac{J_{ij}}{|J_{ij}|} v_{jn}) dZ_{i,j}^{(3)} + (u_{in} - \frac{J_{ij}}{|J_{i,j}|} u_{jn}) dZ_{i,j}^{(3)*}  \Big),
\eea
\bea\label{u_final}
    \frac{d u_{nm}}{dt} &= \big[-(\gamma - \sum_j |J_{nj}|) u_{nm} +\G \left(\Kr{n}{m}+v_{nm}\right) +\G v_{mn}\big] \\
    &-\eta\big[u_{nn}\left(\Kr{n}{m}+v_{nm}\right)+2u_{nm}v_{nn} + \alpha^2_n\left(\Kr{n}{m}+v_{nm}\right)+2\abss{\alpha_n}u_{nm}\big] \\
    &-\eta\big(u_{mm}v_{mn}+2u_{nm}v_{mm} + \alpha^2_m v_{mn}+2\abss{\alpha_m}u_{nm}\big)\\
    &-\sum_j(\gamma - \sum_i |J_{in}|)\left(u_{jn}v_{jm}+u_{jm}v_{jn}\right)-4\sum_j\eta |\alpha_{j}|^2\left(u_{jn}v_{jm}+u_{jm}v_{jn}\right)\\
    &+ \sum_i \sum_{j = 1}^{i-1} (v_{in}-v_{jn} \frac{J_{ij}}{|J_{ij}|}) (u_{im}-u_{jm} \frac{J_{ij}}{|J_{ij}|}) + \sum_i \sum_{j = 1}^{i-1} (v_{im}-v_{jm} \frac{J_{ij}}{|J_{ij}|}) (u_{in}-u_{jn} \frac{J_{ij}}{|J_{ij}|}) \\ 
    & -\frac{1}{2} \sum_j u_{nm} (|J_{mj}| + |J_{nj}|) + \frac{1}{2} \sum_j (u_{nj} J_{mj} + u_{mj} J_{nj} +2\sum_{j}\sqrt{\eta}\left(v_{jn}v_{jm}  dZ_{j,j}^{(2)}+u_{jn}u_{jm}  dZ_{j,j}^{(2)\ast}\right), \\
\eea

\bea\label{v_final}
    \frac{d v_{nm}}{dt}&=\big[- (\gamma - \sum_j |J_{nj}|) v_{nm} +\G u_{nm}^{\ast}+G u_{nm}\big] -\eta\big(u_{nn}^{\ast}u_{nm}+2v_{nn}v_{nm} +2\abss{\alpha_n}v_{nm}+\calphas_n u_{nm}\big)\\
    &-\eta \big(u_{mm}u_{nm}^{\ast}+2v_{mm}v_{nm} +2\abss{\alpha_m}v_{nm}+\alpha^2_m u_{nm}^{\ast}\big)\\
    &-\sum_j(\gamma - \sum_i |J_{in}|)\left(v_{nj}v_{jm}+u_{jn}^{\ast}u_{jm}\right)-4\sum_j\eta\abs{\alpha_{j}}^2\left(v_{nj}v_{jm}+u_{jn}^{\ast}u_{jm}\right)\\
    &+ \sum_i \sum_{j=1}^{i-1} (u_{in}^* - \frac{J_{ij}}{|J_{ij}|} u_{nj}^*) (u_{mi} - \frac{J_{ij}}{|J_{ij}|} u_{mj}^* ) + \sum_i \sum_{j=1}^{i-1} (v_{mi}^* - \frac{J_{ij}}{|J_{ij}|} v_{mj}^*) (v_{ni} - \frac{J_{ij}}{|J_{ij}|} v_{nj}^* )\\
    & - \frac{1}{2} \sum_j v_{nm} |J_{mj}| + \frac{1}{2} \sum_{nj} v_{nj} J_{mj} +2\sum_j\sqrt{\eta}\left(u_{jn}^{\ast}v_{jm} dZ_{j,j}^{(2)}+u_{jm}v_{jn}^{\ast} dZ_{j,j}^{(2)\ast}\right) . \\
\eea
In equation~\eqref{Field_dynamics_equation} we recover the mean-field equations for a system of coupled optical parametric oscillators and in particular the Ising term $+ \frac{1}{2} \sum_j J_{jn} \alpha_j$.  

\section{Threshold value for the pump intensity}
 When pumped by an external drive above the oscillation threshold $G_{\rm th}$, an optical parametric oscillator undergoes a bifurcation instability forcing the phase of the modes to be either $0$ or $\pi$ with respect to the reference phase imposed by the pump. These two states encode the up and down spin configurations of a classical Ising spin. 
The pump threshold $G_{\rm th}$ is defined as the smallest value of the pump $G$ for which there is at least an amplified eigen-mode of the linear system defined by $J_{ij}$.
To compute the threshold value for the pump intensity $G$ we shall start from the mean-field equations. These equations are found from Eq.~\eqref{Field_dynamics_equation} for the dynamics in the Gaussian approximation by setting to zero the second-order correlation function $\bm u$, $\bm v$ and the noise terms $\bm Z$.  
We thus have: 
\bea\label{Mean-field equation}
   \frac{d\alpha_n}{dt} = &\big[ - \frac{\gamma}{2} \alpha_n + \G \calpha_n -  \eta\left(\abss{\alpha_n}\alpha_n\right) \big] + \frac{1}{2} \sum_{j} J_{nj} \alpha_j  .
\eea 
Let us assume that at the initial time $t = 0$ the system is in the vacuum state. In the early stages of the dynamics the mean value $\alpha = \langle
\hat a \rangle$ is expected to be such that $|\alpha_n (t=0)| \ll 1$ for every mode. In this regime, we can neglect nonlinearities in the equations of motion. By considering only the dynamics of the real part of ${\bm \alpha}$, we have: 
\begin{equation}\label{eq:Mean-field-zero-non-lin}
\frac{d \bm \alpha}{d t}\simeq \left(G-\frac{\gamma}{2}\right) \bm \alpha + \frac{1}{2} \bm J \bm \alpha, 
\end{equation}
The coupling matrix is diagonalized as $\bm J = \bm U^\dagger \bm D \bm U$, where $\bm D = {\rm diag}(\lambda_1,.., \lambda_N)$ and $\lambda_j$ are the real eigenvalues of $\mathbf{J}$ since $\bm J$ is symmetric. In the basis $\bm \beta = \bm U \bm \alpha$ Eq.~\eqref{eq:Mean-field-zero-non-lin} becomes: 
\begin{equation}
\frac{d}{d t} \boldsymbol{\beta}=\left(G-\frac{\gamma}{2}\right) \boldsymbol{\beta}+\frac{1}{2}\mathbf{D} \boldsymbol{\beta},
\end{equation}
which is element by element
\begin{equation}
\frac{d \beta_n}{d t}=\left(G-\frac{\gamma - \lambda_n}{2}\right) \beta_n. 
\end{equation}
The $n$-th eigenmode is thus amplified (above threshold) if $G > (\gamma -\lambda_n)/2$. Hence the threshold pump $G_{\rm th}$ is obtained by considering the largest eigenvalue $\lambda_{\text{max}} = \text{max}_n\{\lambda_n\}$, namely, 
\begin{equation}
G_{\rm th} = \frac{\gamma - \lambda_{\text{max}}}{2}.
\end{equation}

\section{Supplementary data}
\begin{figure}[t!]
\includegraphics[width = 0.7\hsize]{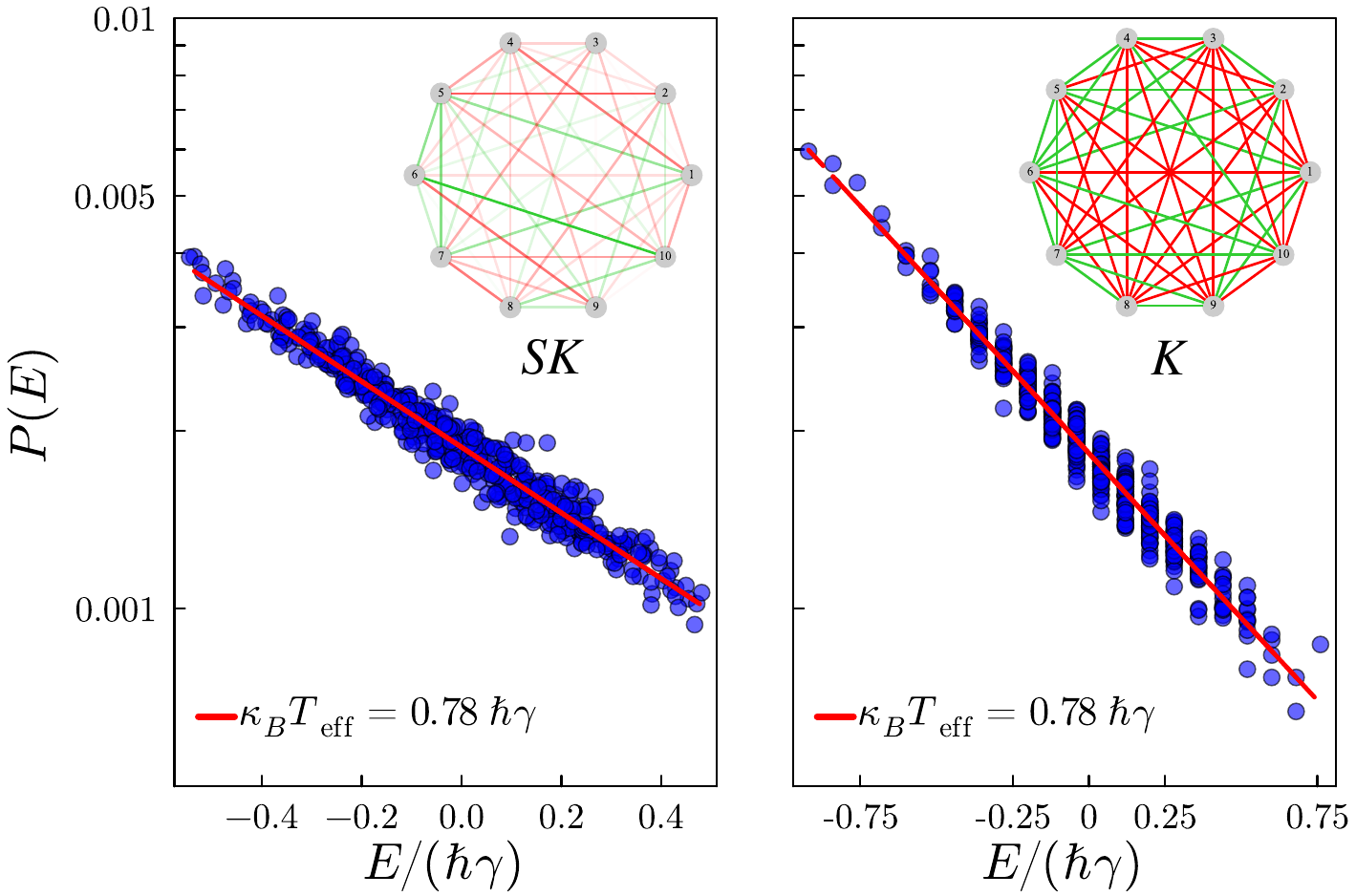}
\caption{Left panel: Plot of the probability distribution $P(E)$ in logarithmic scale of spin configurations with energy $E$ for a system with $N = 10$ modes and coupled via a SK graph (see main text).
Right panel: Same quantity, but for a K graph.  For both graphs, the scaling law $P \propto e^{-E / (k_B T_{{\rm eff}})}$ is observed, revealing an emergent thermal equilibrium with an effective temperature $T_{\rm eff}$. Other parameters: $G/G_{\rm th} = 0.01$, $\eta/\gamma = 1/10$ and $200,000$ time samples taken from a single quantum trajectory evolved for a total time of $t_{\rm max} = 20,000 \, \gamma^{-1}$.   }
\label{fig:double_fit_G_Gth_0_01}
\end{figure}

\begin{figure}[h!]
\includegraphics[width = 0.49\hsize]{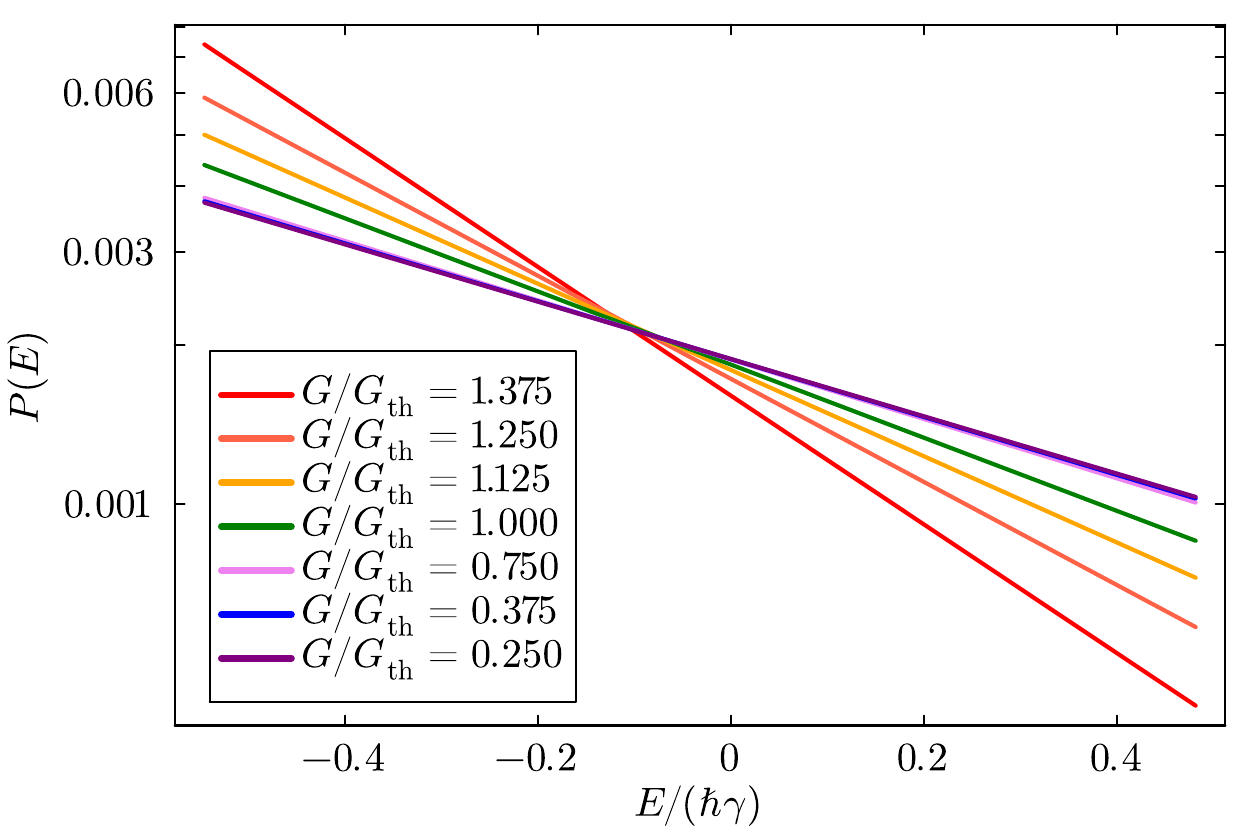}
\includegraphics[width = 0.49\hsize]{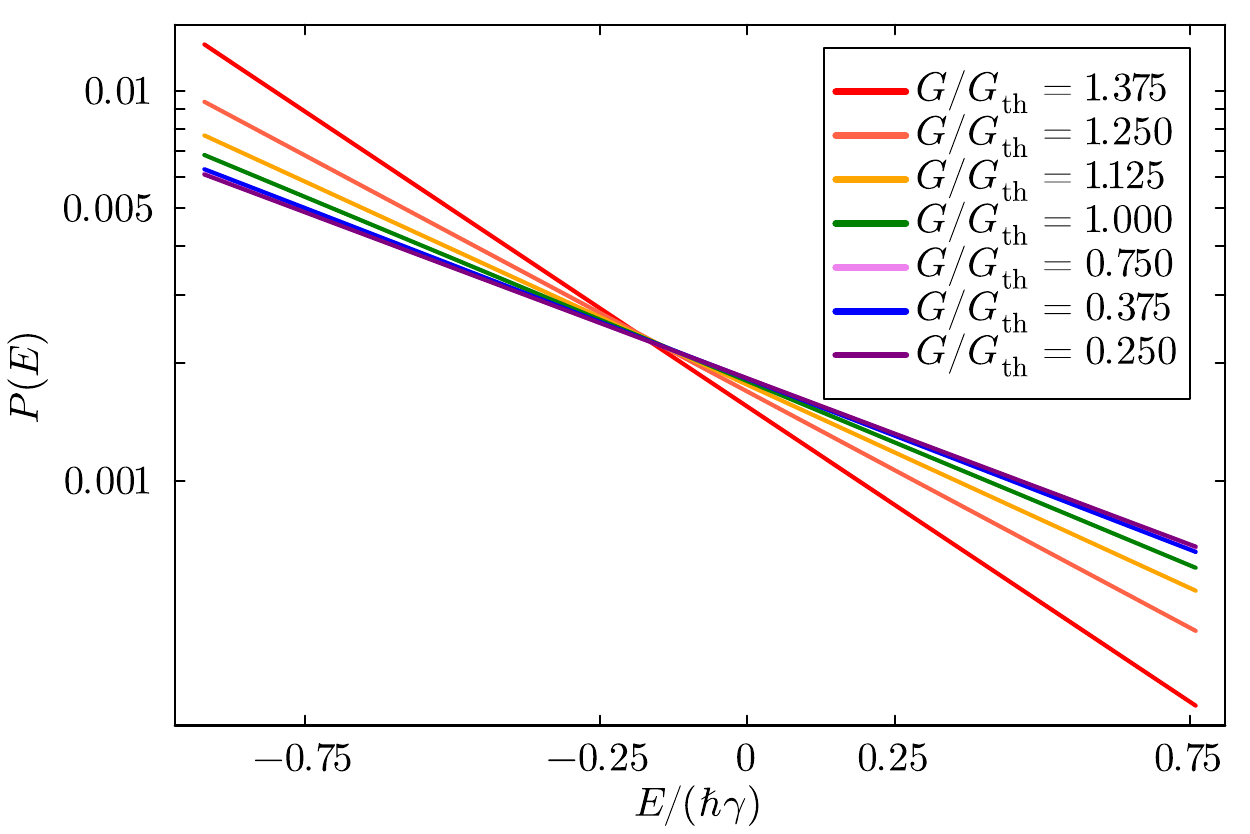}
\caption{Plot of the fitted Boltzmann distributions in logarithmic scale for different values of pump intensity $G/G_{\rm th}$ of the $10$-mode system coupled via (Left) SK graph and (Right) K graph. The lines exhibit an increasing slope for higher values of the pump intensity $G$. Parameters: $\eta/\gamma = 1/10$. Total number of samples per point: $200,000$ ($t_{\rm max} = 20,000 \ \gamma^{-1}$).}
\label{fig:multiple_crossing}
\end{figure}

In this section, we include some additional plots with different parameters with respect to the main text.
Figure~\ref{fig:double_fit_G_Gth_0_01} shows the probability distribution $P(E)$ for the two graphs SK and K considered in the main text, but here for a driving $G/G_{\rm th} = 0.01$. This value gives a Boltzmann distribution with higher temperature compared to the one found in the main text with larger pump amplitude. The fitted Boltzmann distribution for different pump amplitude values is shown in Fig~\ref{fig:multiple_crossing}. This analysis corroborates the findings presented in the main text. In particular, we observe that the Boltzmann distribution at different pump strengths overlap within a narrow energy range, and no significant change of temperature is found for low pump values.

\end{document}